# The local electronic structure of α-Li$_3$N

T. T. Fister[1,2], G. T. Seidler[1,*], E. L. Shirley[3], F. D. Vila[1], J. J. Rehr[1], K. P. Nagle[1], J. C. Linehan[4], J. O. Cross[5]

[1]Physics Department, University of Washington, Seattle, WA 98195
[2]Materials Science Division, Argonne National Labs, Argonne, IL 60439
[3]National Institute of Standards and Technology, Gaithersburg, MD 20899-8441
[4]Pacific Northwest National Labs, Richland, WA 99352
[5]Advanced Photon Source, Argonne National Labs, Argonne, IL 60637

(*) seidler@phys.washington.edu

# ABSTRACT

New theoretical and experimental investigation of the occupied and unoccupied local electronic density of states (DOS) are reported for $\alpha$-Li$_3$N. Band structure and density functional theory calculations confirm the absence of covalent bonding character. However, real-space full-multiple-scattering (RSFMS) calculations of the occupied local DOS finds less extreme nominal valences than have previously been proposed. Nonresonant inelastic x-ray scattering (NRIXS), RSFMS calculations, and calculations based on the Bethe-Salpeter equation are used to characterize the unoccupied electronic final states local to both the Li and N sites. There is good agreement between experiment and theory. Throughout the Li 1$s$ near-edge region, both experiment and theory find strong similarities in the $s$- and $p$-type components of the unoccupied local final density of states projected onto an orbital angular momentum basis ($l$-DOS). An unexpected, significant correspondence exists between the near-edge spectra for the Li 1$s$ and N 1$s$ initial states. We argue that both spectra are sampling essentially the same final density of states due to the combination of long core-hole lifetimes, long photoelectron lifetimes, and the fact that orbital angular momentum is the same for all relevant initial states. Such considerations may be generically applicable for low atomic number compounds.

## I. INTRODUCTION

Solid-state lithium nitrides and their immediate derivatives demonstrate a rich diversity[1] of bonding characteristics and crystal structures which has in turn led to both fundamental interest and a considerable range of potential applications. For example, α-$Li_3N$ serves as the starting point for potential hydrogen storage methods due its high theoretical $H_2$ capacity[2] and also is a component in the synthesis of nanophase GaN.[3] In addition, its related transition-metal (M) substituted compounds[4] $Li_{3-x}M_xN$, are excellent ionic conductors (as is α-Li3N itself[1,5]) and have been proposed as constituents of improved electrodes in Li-based batteries.[6]

After significant investigation and dispute, the chemical bonding of α-$Li_3N$ is now understood to be near the extreme ionic endpoint of the continuum from covalent to ionic behavior.[1] It is the only alkali-nitride phase which is stable at ambient conditions,[7,8] and it is interesting that several estimates of the nominal classical valences of the constituents are close to $Li^+$ and $N^{3-}$.[1] Consequently, α-$Li_3N$, and possibly its high-pressure phases,[8-10] are believed to provide the only known examples of the $N^{3-}$ ion. $N^{3-}$ is not stable as a free ion, and the mechanism by which the hexagonal bipyramid of neighboring $Li^+$ may stabilize $N^{3-}$ in α-$Li_3N$ has been a matter of some interest.[1,9,11-13] For means of context, we show the crystal structure[14] of α-$Li_3N$ in the top panel of Fig. 1. Note the layered structure and the presence of two crystallographically-distinct Li sites.

Here, we revisit the local electronic structure in α-$Li_3N$. First, using band structure, density-functional theory, and real-space full-multiple-scattering (RSFMS) methods, we consider the occupied electronic states. The former two calculations find strong support for extremely ionic bonding character in α-Li$_3$N. The weak deviations from sphericity in the calculated valence charge distribution around N show no sign of covalent character with the Li nearest neighbors. By common measures, such as integration of valence charge density within nominal ionic radii, the charge transfer from Li to N is at least nearly complete. However, calculation of charge transfer based on RSFMS calculation of the occupied local DOS disagree with the above, finding clear evidence for occupied states just below the Fermi energy at all crystallographic sites. We compare and contrast these results, emphasizing the difficulty in uniquely defining species valence, and suggesting the need for experimental measurement of the occupied local DOS, such as by photoemission spectroscopy.

Second, we report a combined experimental and theoretical investigation of the unoccupied electronic states. Excited state electronic spectroscopies such as x-ray absorption spectroscopy (XAS) and electron energy-loss spectroscopy (EELS) offer a local probe of the unoccupied electronic states and have been widely used to study the chemical bonding of low atomic number materials.[15] However, there has been little XAS or EELS work on α-Li$_3$N, with the few existing studies concentrating on the very near-edge region within less than 10 eV of the Li and N *K*-shell binding energies.[16,17] Nonresonant inelastic x-ray scattering (NRIXS) using hard x-rays provides an alternative which allows investigation of both dipole-allowed and dipole-forbidden final states while also avoiding the problematic surface sensitivity of soft x-ray XAS and the possibility of beam damage in EELS. The core shell contribution to NRIXS is generally called non-resonant x-ray Raman scattering (XRS). The first XRS study of the N *K*-edge in several phases of Li$_3$N has recently[9] been reported.

Here, we present the first measurements of the momentum transfer (*q*) dependence of the XRS from the Li 1*s* initial state and also present new measurements over a wider energy range of the effectively dipole-limit XRS from N 1*s* initial states. Our measurements of the Li 1*s* XRS allow experimental determination of the local final density of states projected onto an orbital angular momentum basis (i.e., the *l*-DOS) at the Li sites. All measurements show good agreement with new *ab initio* calculations. We find that the ionic nature of α-Li$_3$N requires a qualitative interpretation of *l*-DOS which is rather different than has previously been used in systems with strong covalent bonding. A similar issue arises when discussing the observed similarity of the near-edge structure for the Li and N 1*s* XRS. Such behavior in XAS studies is often taken as a signature of covalent bonding. However, here we find that it is a generic consequence of the strong multiple scattering limit for low-energy photoelectrons in cases when all core-hole lifetimes are much longer than the photoelectron lifetime and when all initial states have the same orbital angular momentum. Such conditions will in general hold for compounds of low atomic number elements.

## II. THEORY

### A. Nonresonant inelastic x-ray scattering

The double differential cross-section for NRIXS is proportional to the dynamic structure factor,

$$S(\vec{q},\omega) = \left| \langle f | \exp(i\vec{q}\cdot\vec{r}) | 0 \rangle \right|^2 \delta(\hbar\omega - E_f + E_0) \qquad (1)$$

where $\vec{q}$ is the momentum transfer, $\hbar\omega$ is the energy transfer, and $E_0$ ($E_f$) is the energy of the initial (final) state. The contribution to $S(\vec{q},\omega)$ due to excitations from core levels is commonly called nonresonant x-ray Raman scattering (XRS). Beyond the plasmon frequency and when in the dipole-scattering (i.e. low-$q$) limit, the XRS contribution to $S(\vec{q},\omega)$ becomes proportional to the absorption coefficient measured in XAS, albeit with bulk-like sensitivity even for low-energy edges due to the large incident and scattering photon energy.[18,19]

Non-dipole selection rules apply for $S(\vec{q},\omega)$ at higher $q$,[20,21] allowing investigation of the complete range of symmetries of final states at the probe atom.[21-26] For powder samples, the XRS contribution to $S(q,\omega)$ may be recast[21] as

$$S_{core}(q,\omega) = \sum_{l}(2l+1)|M_l(q,\omega)|^2 \rho_l(\omega). \tag{2}$$

In Eq. 2, the projection of the local final density of states onto an orbital angular momentum basis (*i.e.*, the *l*-DOS) is represented by $\rho_l$. The atomic-like $M_l$ coefficients may be readily calculated.[21] At each energy, Eq. 2 should therefore be viewed as a set of linear equations (one for each experimental $q$) involving experimental measurements of $S(q,\omega)$, the known $M_l$ coefficients, and the unknown $\rho_l(\omega)$. As such, it gives a route for experimental determination of $\rho_l(\omega)$ from XRS measurements.[19,23,24,26] Here, we use this formalism to determine both the *s*-type ($\rho_0$) and *p*-type ($\rho_1$) *l*-DOS for Li in α-Li$_3$N. By comparison, XAS or low-$q$ EELS measurements for the Li 1*s* initial state will be sensitive only to $\rho_1$ due to the dipole selection rule. The range of potential applications of this new *l*-DOS spectroscopy has recently been discussed in detail.[23]

**B. Calculations of *l*-DOS and S($q,\omega$)**

Our experimental measurements of the excited electronic states are complemented by two independent *ab initio* theoretical treatments. In both cases, the potentials used in the calculations are determined self-consistently *without* any assumption about the valence of the constituent atoms. This is contrary to the some prior band structure calculations[11] for α-Li$_3$N, where an N$^{3-}$ ion was stabilized by use of an artificial potential prior to enforcing self-consistency.

First, a real-space full-multiple-scattering (RSFMS) approach using the FEFF code[27] was used to calculate the *l*-DOS and XRS spectra[21] (*i.e.*, $S_{core}(q,\omega)$). The full-multiple-scattering matrix includes atoms up to 8 Å from the central scatterer using self-consistent muffin-tin potentials. For both parts of the calculation, the spectra are converged with respect to the number of atoms and the number of angular momentum states, which was limited to $l = 2$. Finally, we include the effects of the core-hole lifetime and the photoelectron's self-energy in the final spectra. Note that independent calculations for the planar and out-of-plane Li atoms are required for comparison with the orientationally-averaged data.

Second, XRS spectra were calculated using the Bethe-Salpeter equation (BSE), which is the equation of motion for an interacting electron-hole pair produced by core-electron excitation. We use the method described in Soininen, *et al.*,[28] but with certain modifications. First, a model self-energy is used to simulate electron lifetime broadening of spectral features.[29] The screening effects on the electron-core hole interaction and the transition matrix elements are evaluated using methods that have been described in detail elsewhere.[30] Multiplet effects (related here only to spin degrees of freedom) are also included.[31] We use norm-conserving pseudopotentials of the Hamann-Schlüter-Chiang type[32] with Vanderbilt-type cut-off functions[33] and separable projectors.[34] Other details of the calculation method follow prior work.[28] All results are well-converged with respect to Brillouin-zone sampling, with enough unoccupied bands included to calculate screening effects and spectra within and beyond the displayed region. Because of the three-fold symmetry with respect to the z-axis, angular averaging over the direction of $\vec{q}$ was achieved using a momentum along the Cartesian (1,1,1) direction. The final theoretical results are convolved with experimental broadening (1.3 eV full-width half-maximum).

## C. Calculations of electron density and occupied electronic states

The spatial distribution of valence electron density has been calculated in two independent ways. First, it is determined from superposition of the density associated with the bands calculated as a preliminary step in the BSE approach described above. Second, we have performed density-functional theory (DFT) calculations on the experimentally determined structure of $\alpha$-$Li_3N$ using ultrasoft pseudopotentials and a plane-wave basis set. These calculations were carried out with the Vienna *Ab-initio* Simulation Package[35] (VASP) using the PBE functional[36] and an energy cutoff of 350 eV. The Bader decomposition[37] of the electron density uses the algorithm introduced by Henkelman *et al.*[38]

The occupied local DOS on each crystallographic site was calculated with the same RSFMS approach as presented in section II.B, except that no core-hole was included in the calculations. The valence electron density and unoccupied local DOS calculations provide complementary benefits and perspectives as regards the question of charge transfer, as we discuss below.

### III. EXPERIMENTAL

All NRIXS measurements were performed with the lower energy resolution inelastic x-ray scattering (LERIX) spectrometer at the PNC/XOR 20-ID beamline of the Advanced Photon Source.[39] Spectral normalization, subtraction of the valence Compton background, and other data processing issues follow methods reported elsewhere.[22,40] The energy resolution was 1.3 eV. A dense 5 mm thick pellet of commercially-prepared α-Li$_3$N powder (99 % purity) was measured in a transmission geometry. Note that the 5 mm thickness is chosen to match the penetration length of α-Li$_3$N at 10 keV. With such a large penetration length, our measurements are inherently bulk-like with no sensitivity to surface contamination. While commercial Li$_3$N typically contains a small fraction of metastable β-Li$_3$N impurity, the bulk sensitivity of XRS ensures that the measured signal is dominated by the desired α-Li$_3$N phase.

The incident flux of photons was about 5 10$^{12}$ s$^{-1}$. The count rates at the edges and in the underlying valence Compton background are strong functions of $q$. At $q = 2.4$ Å$^{-1}$ we find 6100 and 230 counts/s at the Li and N $K$-edge step on valence Compton scattering backgrounds of 17000 and 220 counts/s, respectively. Increasing $q$ to 9.8 Å$^{-1}$ resulted in 2100 and 1300 counts/s at the Li and N $K$-edge step on valence Compton scattering backgrounds of 1100 and 31000 counts/s, respectively. Following these measurements we found a significant air leak into the He flight path (outside the sample space) of the instrument. This resulted in a decrease in all count rates by a factor of about 3.5 compared to present LERIX performance, but otherwise has no consequences for the results presented here. The sample was enclosed in a flowing He environment during measurements. The XRS spectra show no evidence of evolution with measuring time, minimizing concerns about beam damage.

## IV. RESULTS AND DISCUSSION

### IV. A. Occupied electronic states

The highly ionic nature of $\alpha$-Li$_3$N plays an important role in the interpretation of NRIXS results (below). As there has been significant progress in computational methods in the several years since the spatial distribution of electron density was last directly investigated for this system,[11-13] it is worthwhile to revisit this issue. In the middle and bottom panels of Fig. 1, we show the calculated valence electron density in the Li$_2$N basal plane (001) and the normal plane ($1\bar{1}0$) based on the new band-structure calculations. The contours are on a logarithmic scale, as explained in the caption. The valence charge density is nearly spherical around the N sites. Note that the first deviations from sphericity show a weak excess of charge density in the N-N nearest neighbor direction, not the N-Li direction, so that even at low valence charge densities there is no signature of covalent character. The valence charge density is essentially zero at the Li nuclei, and decreases monotonically and nearly exponentially after a local maximum near the N nuclei.

This can be seen more clearly in the top panel of Fig. 2 which shows the directionally-averaged valence charge density as a function of distance from the three crystallographic sites. The vertical dashed line in the figure is at the Li-N nearest neighbor distance in the Li$_2$N sheets. The bottom panel of the figure shows the integrated valence charge as a function of distance from the three crystallographic sites, together with the total Li-site charge. These results strongly support a very ionic picture for the chemical bonding in α–Li$_3$N, and are very similar to prior band-structure based results.[11,13]

In Fig. 3, we show the α–Li$_3$N valence electron density surface for a cutoff of 0.07 electrons/Å$^3$, calculated with the DFT/VASP approach. The blue and orange spheres indicate the positions of the N and Li atoms, respectively, and the cell is oriented to present the Li$_2$N basal plane on the top. The electron density around the N atoms is very close to spherically symmetric, even in regions close to the Li atoms. Small deviations are visible on the basal plane, and have the same characteristics as seen in the band-structure calculations. When the electron density distribution is analyzed using the "Atoms in Molecule" scheme of Bader,[37] we find that the dividing surfaces are pushed into the Li atom positions, resulting in a highly ionic distribution. This can be seen for the surface of the contour depicted in the figure, which lies very close to the Li atoms, but does not show a spherical component close to them. The empirical charges estimated from these dividing surfaces reveal fully ionized Li atoms and therefore N atoms carrying three extra charges, and the general behavior of the dividing surface reinforces that α-Li$_3$N has little covalent character.

However, it is important to remember that the valence charge of a particular species in a solid is not itself well-defined. Any correct approach to the question of charge transfer must explicitly address a quantum mechanical (*i.e.* experimental) observable, and the final estimate of charge transfer may indeed depend on the choice of observable. In Fig. 4, we present RSFMS calculations of the occupied local DOS for cells surrounding each of the three crystallographic sites. The key feature of these results is not just the presence of large N $2p$ DOS just below $E_F$, but also the presence of significant DOS at the similar energies for both Li sites. The occupied charge counts are quite different from what was qualitatively inferred above from the band structure or DFT/VASP calculations or from prior band structure calculations[11-13] because they refer to the contributions of the global DOS projected onto the given cells. The RSFMS calculations provide nominal valences of $N^{-0.9}$ and $Li^{+0.3}$, with slightly higher ionization of the out of plane Li site and slightly lower ionization at the in-plane Li site.

This significantly lower estimate of charge transfer may help resolve a dilemma which has been noted in comparing the electronic structures of α-, β-, and γ-$Li_3N$.[9] In Lazicki, *et al.*,[9] it was proposed that high-pressure γ-$Li_3N$ must be significantly more ionic than α- or β-$Li_3N$ due to a large increase in the band gap. However, this is difficult to rationalize if α-$Li_3N$ already has complete ionization of all species in addition to the absence of covalent bonding character. In future work, it would be interesting to consider how strongly the nominal valence of the various species depends on the definition of the local DOS. For example, it would clearly be valuable to have direct experimental measurement of the density and symmetry of occupied states, such as by angle-resolved photoemission spectroscopy

**IV.B. Excited electronic states**

With the above reinforcement that chemical bonding in α–$Li_3N$ does indeed have very little covalent character, we proceed to present and discuss the new experimental results. In the top of Fig. 5, we show Li 1s XRS spectra for $q = 0.8, 2.4, 3.9, 5.3, 6.6, 7.7, 8.6, 9.3, 9.8$ and $10.1$ Å$^{-1}$. Counts are integrated for 9 s at each point. The corresponding BSE calculations are shown in the lower panel. To aid comparison, the energy loss spectra have been normalized by their integrated intensity over the first 30 eV and are vertically displaced between successive $q$. Several features in the near-edge spectrum are labeled ($a - f$ in Fig. 5) to highlight the agreement between theory and experiment. Note that each of these features exhibit modest $q$-dependence, with $a, b,$ and $e$ decreasing in relative amplitude with $q$ while $c, d,$ and $f$ either increase or only appear at higher momentum transfers. Each feature corresponds to a peak in the unoccupied final density of states local to each Li atom, albeit with each possessing differing local symmetry. Such $q$-dependence in localized features can be interpreted as changes in the relative weighting of the different $l$-DOS components.[21,23] We note that the XRS spectrum for the lowest measured $q$ disagrees significantly with recent XAS results.[16] The bulk sensitivity of XRS, the good agreement between theory and experiment, and the preferential surface sensitivity of XAS with soft x-rays give us confidence in the present results.

On calculating $M_l(q,\omega)$ for the Li 1s initial state,[21,40] we find that only the s- and p-DOS ($l = 0$ and 1 respectively) significantly contribute to $S(q,\omega)$. After normalizing the data shown in Fig. 5 to absolute units, we extract $\rho_0$ and $\rho_1$ at each energy point in the near-edge regime using a least squares fit to the model given by the first two terms in Eq. 2.[23,40] The $M_l$ coefficients were found to be nearly identical for the two crystallographically independent Li sites; hence, only the stoichiometrically-averaged $l$-DOS from both sites can be obtained by inverting the experimental results via Eq. 2.

These results are shown in Fig. 6a. The RSFMS calculations for the *p*-DOS and *s*-DOS are shown in Figs. 6b and 6c, respectively. The agreement between the experimental and calculated *l*-DOS is relatively good, with discrepancies mainly in the relative amplitude of features. Such disagreements are to be expected in the present RSFMS approach due to the details of the treatment of the electron-core hole interaction and perhaps especially due to our decision to not force the effective radius of the N to grow and of the Li to shrink, *e.g.*, as is needed to artificially force agreement with the expectation of a large nominal N ionic radius, such as is seen in Figs. 1, 2, and 3. Future RSFMS work based on a full-potential calculation should remedy this deficiency.

Some aspects of the experimental and theoretical *l*-DOS deserve special attention. Except for an overall scale factor, the experimentally-determined *s*- and *p*-DOS are quite similar for the first 15 eV. The theoretical *l*-DOS calculations show similar agreement, but clarify that this phenomenon occurs independently for each of the two Li sites. Commonality of isolated features in the *l*-DOS can be taken as a significant fingerprint for covalency, such as from a shared DOS from a hybridized antibonding molecular orbital.[23,24] Given the broad range of energy over which the *s*- and *p*-DOS are similar and given the absence of covalent character in α-Li$_3$N, such an explanation is incorrect here. Instead, the general agreement in structure of the *s*- and *p*-DOS is a simple consequence of the overall symmetry of the crystal structure, *i.e.*, of the potential seen by the photoelectron. A low-energy photoelectron starting at the Li in-plane (interplanar) site will sample a potential with 3-fold (6-fold) rotational symmetry. A strong *s-p* hybridization of final state wavefunctions naturally follows for low-energy photoelectrons, as the long core-hole and photoelectron lifetimes will enforce a strong multiple scattering limit. Such behavior should be generic in strongly ionic materials: features at all energies in the near-edge region will be common across the different $l$-DOS components, with the ratio of amplitudes of $l$-DOS components determined by the details of the local symmetry of the excitation site.

We show experimental and theoretical results for the N 1$s$ XRS in Fig. 7. Counts are integrated for 5 s at each incident photon energy, and each point represents the weighted average of energy loss spectra collected at $q$ = 2.4, 3.9, and 5.3 Å$^{-1}$. The $q$ range for LERIX is insufficient for $l$-DOS determination at the N sites and the results are dominated by transitions to $p$-type final states (*i.e.*, dipole-limited transitions); this is due to the smaller size of the N 1$s$ initial state as compared to the Li 1s initial state. Note that the very near-edge spectrum is in fair agreement with prior XRS measurement[9] having poorer counting statistics but is in disagreement with prior EELS measurement,[17] which we attribute to likely electron beam damage. The agreement with the RSFMS and BSE calculations (solid lines in Fig. 6) is quite good.

In Table 1 we show a comparison of the peak positions for the Li and N 1$s$ XRS spectra, *i.e*., from Figs. 5 and 7. The agreement in the energy spacing between the first and second features is quite good, and the simple fact of a significant feature at the edge and only one other broad feature in the first 10 eV should not be overlooked. Continuing to higher energies, the absence of peak *(c)* in the N XRS spectrum is expected in this picture: that peak is due to $s$-type final states which are inaccessible at the N $K$-edge for our available $q$ range when using ~10 keV incident photons. The modest energy shifts of peaks *(d)* and *(e)* may be due to differences in the self energy effects for the two different species.

The peak matching of the Li and N 1*s* near-edge spectra indicates that they result from different sampling of a shared, underlying DOS. This is most often seen in XAS spectra for systems with strong covalent bonding, where strong, localized, multiple-scattering resonances (*i.e.*, antibonding orbitals) necessarily span neighboring crystallographic sites and result in peak matching in the near-edge spectra. However, we propose that the same behavior can occur in ionic systems, in the absence of such localized resonances, when the core-hole and photoelectron lifetimes are large.

Most generally, in the limit of long lifetimes the x-ray absorption coefficient $\mu_j(E)$ at a crystallographic site *j* is

$$\mu_j(E) = \sum_k \langle j|d|k,\bar{j}\rangle \delta(E-E_k)\langle k,\bar{j}|d|j\rangle \qquad (3)$$

where *k* is the photoelectron momentum, *d* is the dipole transition operator, $|k,\bar{j}\rangle$ is the photoelectron wavefunction for momentum *k* including final state effects from the core hole at site *j*, and $|j\rangle$ denotes the initial state, localized orbital at site *j*. For ionic compounds, core-hole effects are relatively small in the limit of long lifetimes because the photoelectron wavefunction will be spatially extended. Hence, in this limit,

$$\mu_j(E) \approx \sum_k |\langle k|d|j\rangle|^2 \delta(E-E_k). \qquad (4)$$

As $|f_j\rangle \equiv d|j\rangle$ is simply the dipole-allowed final states accessible from $|j\rangle$,

$$\mu_j(E) \approx \sum_k |\langle k|f_j\rangle|^2 \delta(E-E_k) = \langle f_j|\rho^{global}(E)|f_j\rangle \qquad (5)$$

where $\rho^{global}(E)$ is the (global) density of states for the system. For 1*s* initial states, such as here, the same orbital angular momentum projected component of $\rho^{global}(E)$ will be important for all sites, i.e., the *p*-type component $\rho_1^{global}(E)$. The near-edge spectra are therefore expected to be similar, with differences in the relative amplitude of near-edge features due to the fine spatial details of the expectation integrand in Eq. 5. As an extreme case, for example, note that geometric arrangements (such as three-atom collineations[41]) which strongly favor particular high-order multiple scattering paths will lead to differences in $\mu_j(E)$ between sites involved in such paths and those independent of such paths. Modest differences in the relative locations of peaks may be caused by the site- and final-state-specific nature of the interaction between the core-hole and photoelectron.

Compounds of low *Z* elements will necessarily have long core-hole lifetimes for the 1*s* initial state, and the above argument should be relevant in the near-edge region when the photoelectron lifetime is also relatively large. As photoelectron energy increases this mechanism is cut off by the increase in extrinsic losses and consequent decrease in photoelectron lifetime[42], so that $\left|k,\bar{j}\right\rangle$ and the fine structure then become sensitive to the source site. We therefore anticipate that significant commonality in the XAS, EELS, or dipole-limit XRS near-edge structure at different crystallographic sites may be routinely observed in low *Z* systems when the selected initial states at all sites have the same orbital angular momentum. When outside of the dipole limit, the different admixtures of final state symmetries present in the XRS spectra may result in unique spectral features, such as is the case for feature *(c)* in the Li XRS of the present study.

## V. CONCLUSIONS

In conclusion, we report measurement and *ab initio* calculation of the local electronic structure of α-$Li_3N$. First, band-structure and density functional theory calculations reinforce that chemical bonding in α–$Li_3N$ has little covalent character Interestingly, real-space full multiple scattering calculations of nominal valences based on a quantum mechanical observable, the local final density of occupied states, finds relatively modest charge transfers resulting in $N^{-0.9}$ and $Li^{+0.3}$ ions. This is contrary to more empirical measures of nominal valence, such as from integrating the calculated valence charge density in a selected ionic radius. Second, we find good agreement between theory and both the excited state spectra from the Li and N *K*-edges and also the resulting experimentally-determined *l*-DOS. The strong correspondence between *s*- and *p*-type DOS in the experimental measurement and in the theoretical calculations is explained as a general consequence of the symmetry of the crystal structure, and such hybridization over a wide energy range should be a general phenomenon in ionic crystals of low-Z elements. Third, we observe that the N and Li 1*s* near-edge structures are very similar, indicating that they are sampling a common underlying density of states. We propose a new explanation for such behavior which is based on the long core-hole lifetime for low atomic number systems and the extreme multiple scattering limit for low energy photoelectrons. This mechanism should be active in many other materials, independent of whether their chemical bonding is ionic or covalent.

This research was supported by the U.S. Department of Energy, Basic Energy Science, Office of Science, Contract Nos. DE-FGE03-97ER45628 and W-31-109-ENG-38, Office of Naval Research Grant No. N00014-05-1-0843, and the Summer Research

Institute program at the Pacific Northwest National Laboratory. The operation of Sector 20 PNC-CAT/XOR is supported by the U.S. Department of Energy, Basic Energy Science, Office of Science, Contract No. DE-FG03-97ER45629, the University of Washington, and grants from the Natural Sciences and Engineering Research Council of Canada. The operation of Sector 20 PNC-CAT/XOR is supported by DOE Basic Energy Science, Office of Science, Contract No. DE-FG03-97ER45629, the University of Washington, and grants from the Natural Sciences and Engineering Research Council of Canada. Use of the Advanced Photon Source was supported by the U.S. Department of Energy, Basic Energy Sciences, Office of Science, under Contract DE-AC02-06CH11357. We thank Ed Stern, Warren Pickett, Micah Prange, Joshua Kas, and J. Aleksi Soininen for helpful discussions.

**Figure and Table Captions**

**Fig. 1**: **Top panel**: A ball-and-stick schematic of the crystal structure of α-Li$_3$N. The small blue spheres represent N atoms in the Li$_2$N basal planes, while the larger orange and red spheres represent the in-plane (Li(1)) and out-of-plane (Li(2)) Li sites, respectively. **Middle panel**: The valence electron charge density for α-Li$_3$N in the (001) plane from band structure calculations. **Bottom panel**: The valence electron charge density in the $(1\bar{1}0)$ plane from band structure calculations. The contours in these two panels are logarithmically-spaced, and the central atom for both panels is N. In the middle panel, the outermost triangular contour around Li sites has $\rho_{valence}=10^{-2}$ electrons/Å$^3$ and the contours are spaced by $\Delta \log_{10} \rho_{valence} = 0.2$ with $\rho_{valence}$ again in units of electrons/Å$^3$. In the bottom panel, the outermost roughly square contour around the N site has $\log_{10} \rho_{valence} = -2.2$ and the contour spacing is $\Delta \log_{10} \rho_{valence} = 0.2$ with $\rho_{valence}$ in units of electrons/Å$^3$.

**Fig. 2**: (Top): The directional average of the calculated valence charge density around each crystallographic site in α-Li$_3$N, based on the band structure calculation. The units for the plot are electrons/Å$^3$. (Bottom): The integrated charge as a function of radial distance around each atomic site. The vertical dashed line indicates the nearest-neighbor Li-N distance in the Li$_2$N basal plane.

**Fig. 3**: The valence electron density surface for a cutoff of 0.07 electrons/Å$^3$ for α-Li$_3$N, as determined by DFT calculations; see the text for details. The top surface of the sample is a Li$_2$N basal plane. The blue and orange spheres are N and Li atoms, respectively.

**Fig. 4**: RSFMS calculations of the occupied $l$-DOS for N (left), the in-plane Li site (center), and the out-of-plane Li site (right).

**Fig. 5**: The Li 1$s$ contribution to $S(q,\omega)$ for α-Li$_3$N. Top panel: experimental results stacked by $q$ and normalized to aid comparison. Statistical errors are comparable to the size of the dots used in the figure. Bottom panel: analogous BSE results with additional broadening from the experimental resolution and the self-energy of the photoelectron. The first six notable features are labeled $a - f$ to highlight the agreement between experiment and theory.

**Fig. 6**: (a) The experimental Li $s$- and $p$-DOS ($l = 0$ and 1 respectively) given by inverting the results shown in Fig. 4. Shown by gray dashed lines, the 95 % confidence interval is given by the quality of the least-squares fit to Eq. 2 at each energy point; (b) the total theoretical $p$-DOS is given by the dark curve, while the contributions from the planar and interplanar sites are given by the gray and dashed curves respectively; (c) the analogous figure for the theoretical $s$-DOS.

**Fig. 7**: The N 1$s$ contribution to $S(q,\omega)$ for α-Li$_3$N. Top curve: experimental results averaged between $q = 2.4$ Å$^{-1}$ and 5.3 Å$^{-1}$. Standard uncertainties are comparable to the size of the dots used in the figure. Bottom curve: theoretical results from BSE calculations.

**Table 1:** A comparison of the energies and energy spacings for the principal features observed in the Li and N 1$s$ XRS spectra for α-Li$_3$N, as per Figures 4 and 6. Feature $c$ occurs only in the Li XRS spectra as a consequence of nondipole selection rules at high momentum transfer.

**Figures and Tables**

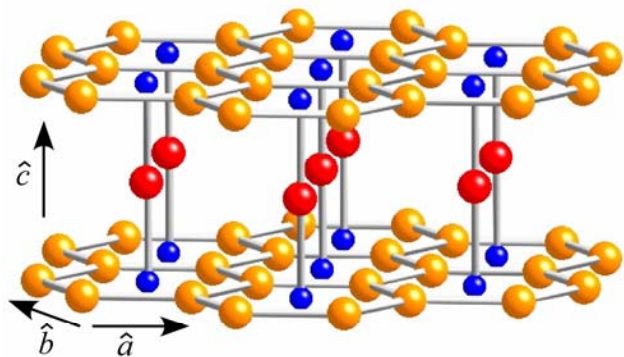

**Figure 1:** T. T. Fister, *et al.*, "The local electronic structure of α-Li$_3$N", submitted, J. Chem. Phys., 2007.

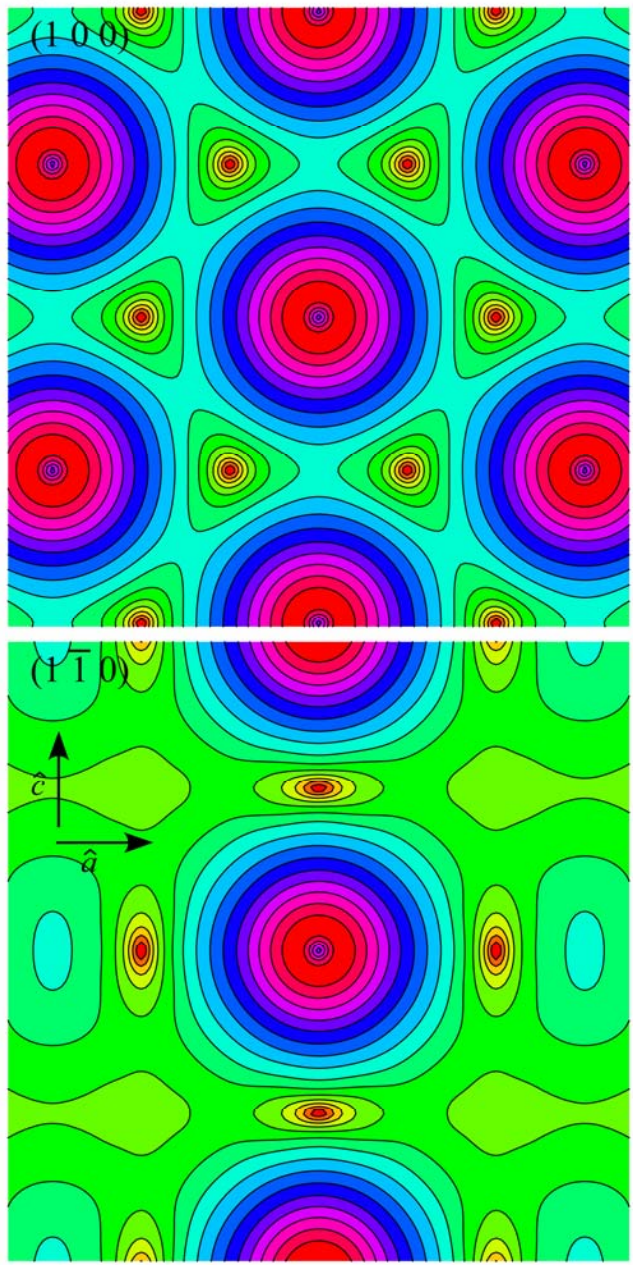

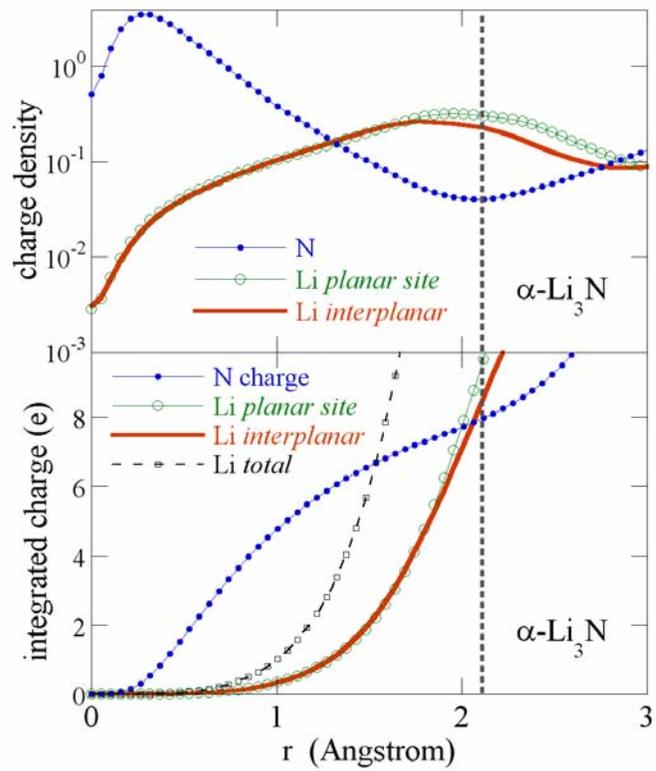

**Figure 2:** T. T. Fister, *et al.*, "The local electronic structure of α-Li$_3$N", submitted, J. Chem. Phys., 2007.

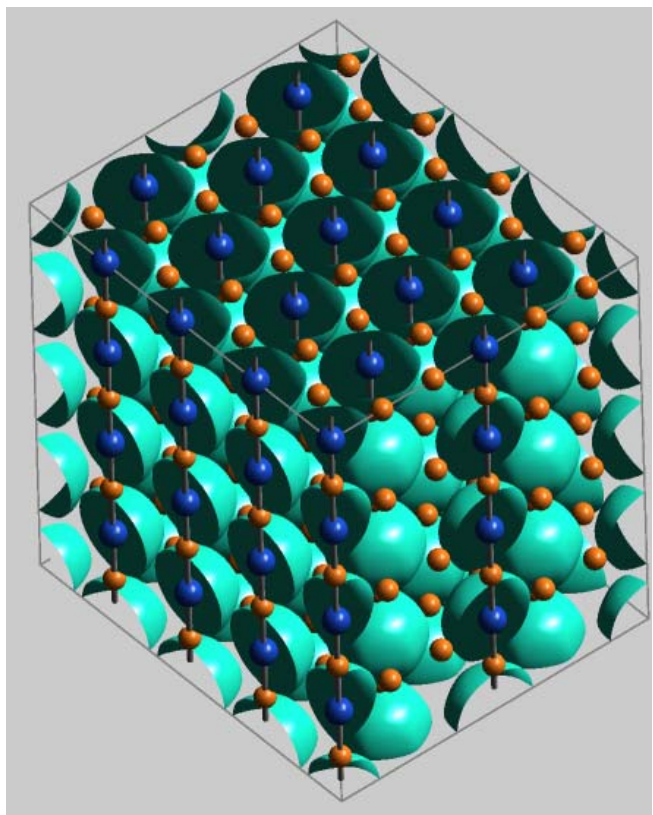

**Figure 3:** T. T. Fister, *et al.*, "The local electronic structure of α-Li$_3$N", submitted, J. Chem. Phys., 2007.

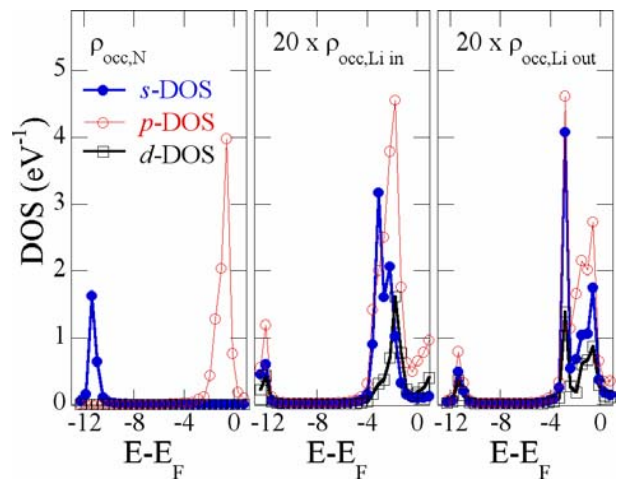

**Figure 4:** T. T. Fister, *et al.*, "The local electronic structure of α-Li$_3$N", submitted, J. Chem. Phys., 2007.

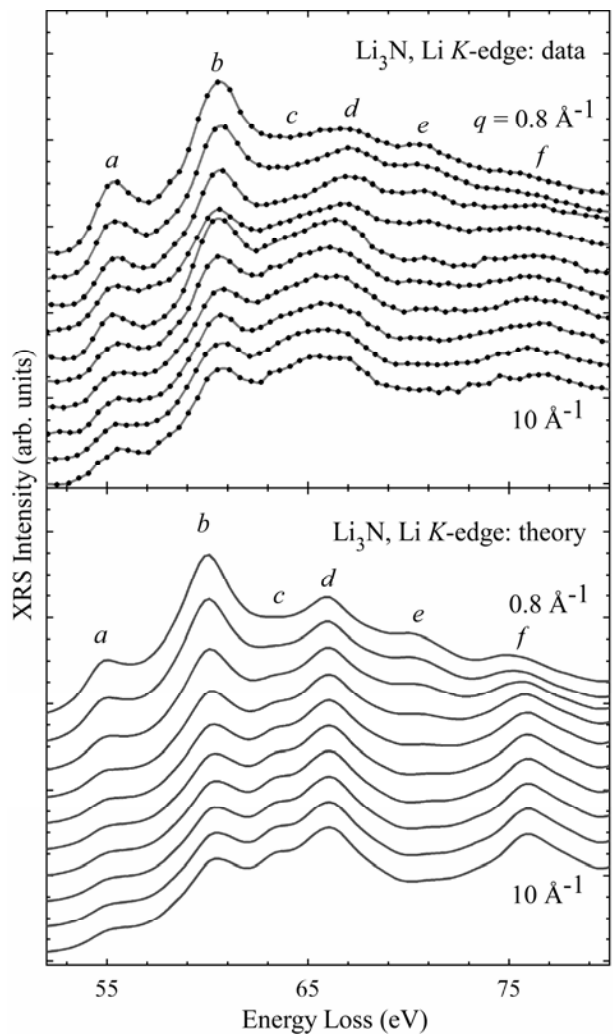

**Figure 5:** T. T. Fister, *et al.*, "The local electronic structure of α-Li3N", submitted, J. Chem. Phys., 2007.

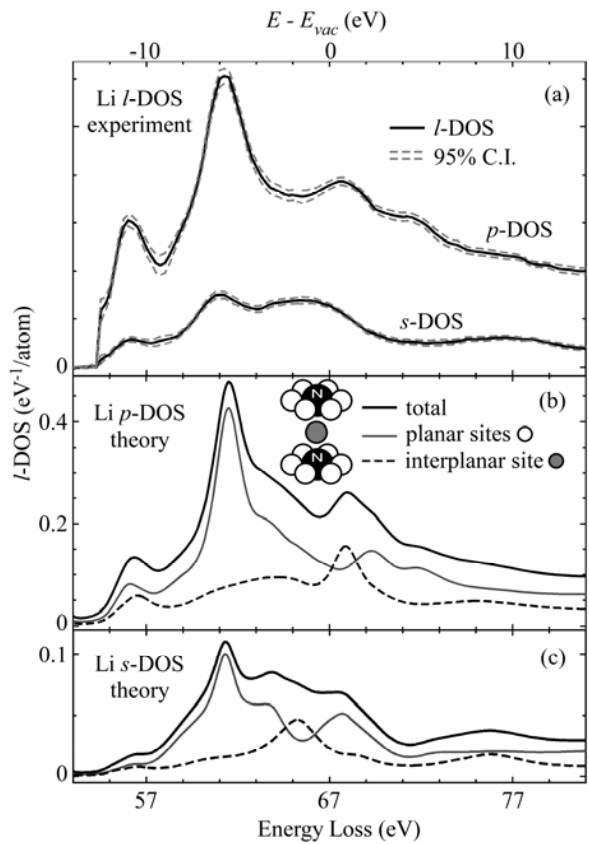

**Figure 6:** T. T. Fister, *et al.*, "The local electronic structure of α-Li$_3$N", submitted, J. Chem. Phys., 2007.

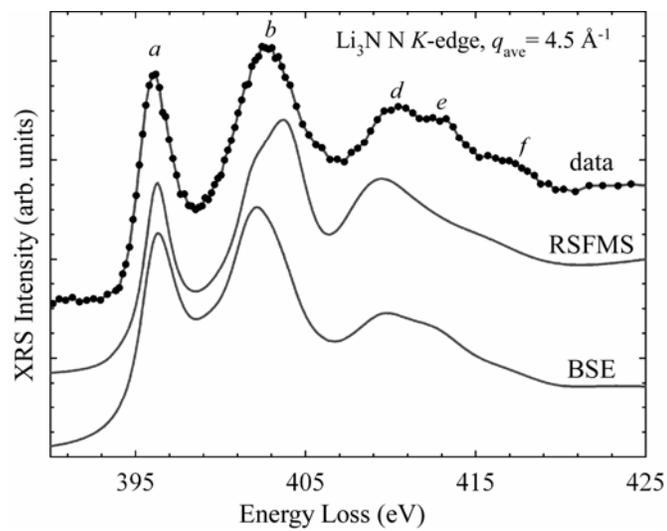

**Figure 7:** T. T. Fister, *et al.*, "The local electronic structure of α-Li$_3$N", submitted, J. Chem. Phys., 2007.

|         | Li 1s   |            | N 1s    |            |
|---------|---------|------------|---------|------------|
| feature | E (eV)  | E-E$_a$ (eV) | E (eV)  | E-E$_a$ (eV) |
| *a*     | 55.3    | 0          | 396.0   | 0          |
| *b*     | 60.6    | 5.3        | 402.6   | 6.6        |
| *c*     | 63.3    | 8.0        | -       | -          |
| *d*     | 67.1    | 11.8       | 410.4   | 14.4       |
| *e*     | 70.7    | 15.4       | 413.3   | 17.3       |
| *f*     | 76.5    | 21.2       | 417.0   | 21.0       |

**Table 1:** T. T. Fister, *et al.*, "The local electronic structure of α-Li$_3$N", submitted, J. Chem. Phys., 2007.